\documentclass[amsmath,amssymb,aps,prl,twocolumn]{revtex4-1}

\usepackage{graphicx}
\usepackage{dcolumn}
\usepackage{bm}
\usepackage{braket}
\usepackage{inputenc}
\usepackage{color}
\usepackage{paralist}
\usepackage{upgreek}
\usepackage[bookmarks = false]{hyperref}

\begin{document}
\title{Experimental Creation of Single Rydberg Excitations via Adiabatic Passage}

\author{Ming-Ti Zhou$^{1,\,2,\,*}$}
\author{Jian-Long Liu$^{1,\,2,\,*}$}
\author{Peng-Fei Sun$^{1,\,2}$}
\author{Zi-Ye~An$^{1,\,2}$}
\author{Jun Li$^{1,\,2}$}
\author{Xiao-Hui Bao$^{1,\,2}$}
\author{Jian-Wei Pan$^{1,\,2}$}

\affiliation{$^1$Hefei National Laboratory for Physical Sciences at Microscale and Department
of Modern Physics, University of Science and Technology of China, Hefei,
Anhui 230026, China}
\affiliation{$^2$CAS Center for Excellence in Quantum Information and Quantum Physics, University of Science and Technology of China, Hefei, Anhui 230026, China}
\affiliation{$^*$These two authors contributed equally to this work.}

\begin{abstract}
In an atomic ensemble, quantum information is typically carried as single collective excitations. It is very advantageous if the creation of single excitations is efficient and robust. Rydberg blockade enables deterministic creation of single excitations via collective Rabi oscillation by precisely controlling the pulse area, being sensitive to many experimental parameters. In this paper, we implement the adiabatic rapid passage technique to the Rydberg excitation process in a mesoscopic atomic ensemble. We make use of a two-photon excitation scheme with an intermediate state off-resonant and sweep the laser frequency of one excitation laser. We find the chirped scheme preserves internal phases of the collective Rydberg excitation and be more robust against variance of laser intensity and frequency detuning.
\end{abstract}

\maketitle

Single collective excitations play an essential role in quantum information processing with atomic ensembles~\cite{Sangouard2011,Barrett2010}. Especially, due to the collectively enhanced interaction, the excitations can be efficiently converted to single photons in well defined modes, thus making them ideally suited for the implementation of quantum light-matter interface~\cite{Duan2001,Kimble2008} for quantum repeater~\cite{Briegel1998} and quantum networks~\cite{Wehner2018}. In these applications, it is of great advantage if the excitation creation process is efficient and robust. Following the initial DLCZ protocol~\cite{Duan2001}, spontaneous Raman scattering becomes a ubiquitous method of creating single collective excitations. Nevertheless, the creation process is probabilistic, for which the excitation probability has to be kept very low to limit the contribution of high-order excitations. Even though the creation process is heralded by the scattered Raman photon enabling scalable extension to multiple sources, the low excitation probability will merely result in low-rate applications~\cite{Sangouard2011} and impose a demanding requirement on long-lifetime storage~\cite{Radnaev2010,Bao2012,Xu2013k,Cho2016,Yang2016}.

To overcome the probabilistic creation issue, one may consider using external deterministic photon sources and storing them in atomic ensembles. In this way, the collective excitations created are deterministic in principle. Nevertheless, the difficulty is shifted to the development of photon sources~\cite{Somaschi2016,Wang2019} and solving the mismatching issues~\cite{Keil2017} such as frequency, bandwidth etc. A more elegant method is to inhibit high-order excitations directly via interactions~\cite{lukin2001dipole,saffman2002creating,Pedersen2009}. Rydberg blockade~\cite{Saffman2010c,Comparat2010} is such a mechanism, which enables creation of a single excitation in a mesoscopic atomic ensemble ($\sim$$\upmu$m). In this case, an ensemble of atoms behaves as a super-atom which undergoes Rabi oscillation between a ground state and a collective Rydberg state~\cite{dudin2012strongly}. By fixing the pulse area as $\pi$, a single excitation in a Rydberg state can be created deterministically in principle~\cite{saffman2002creating,dudin2012observation}. While, the Rabi frequency is collectively enhanced ($\Omega_N=\sqrt{N}\Omega$, with $N$ for the atom number), leading to a difficulty of controlling the pulse area precisely due to the fluctuation of atom number. Moreover, one has also to control other parameters precisely such as laser intensities and frequency detunings.

Adiabatic passage~\cite{Vitanov2001} is a well-known method to realize robust population inversion between two energy levels. For Rydberg atoms, the excitation process typically makes use of a two-photon scheme involving three atomic levels in a ladder configuration~\cite{saffman2002creating}. In this case, the technique of stimulated Raman adiabatic passage (STIRAP)~\cite{Bergmann1998,Vitanov2001,Vitanov2017} suits very well, which achieves population inversion by applying counterintuitive pulses that couple with the top and bottom transitions respectively. This scheme works well for single atoms. While for atomic ensembles, it is theoretically found that STIRAP with on-resonance pulses results in a fully-dephased collective Rydberg excitation~\cite{Petrosyan2013} that can not be retrieved as a single photon, thus it can hardly be useful for the applications that harness collective enhancement. In contrast, the chirped excitation scheme by sweeping the laser frequency provides a promising alternative~\cite{beterov2011deterministic}. In this paper, we experimentally implement this chirped scheme for a mesoscopic atomic ensemble. We find this excitation scheme preserves internal phases of the collective state via observing efficient retrieval of single photons. The chirped scheme is also tested to be more robust than the traditional $\pi$-pulse scheme in terms of sensitivity of laser pulse area and frequency detuning.

\begin{figure*}[t]
\includegraphics[width=1.55\columnwidth]{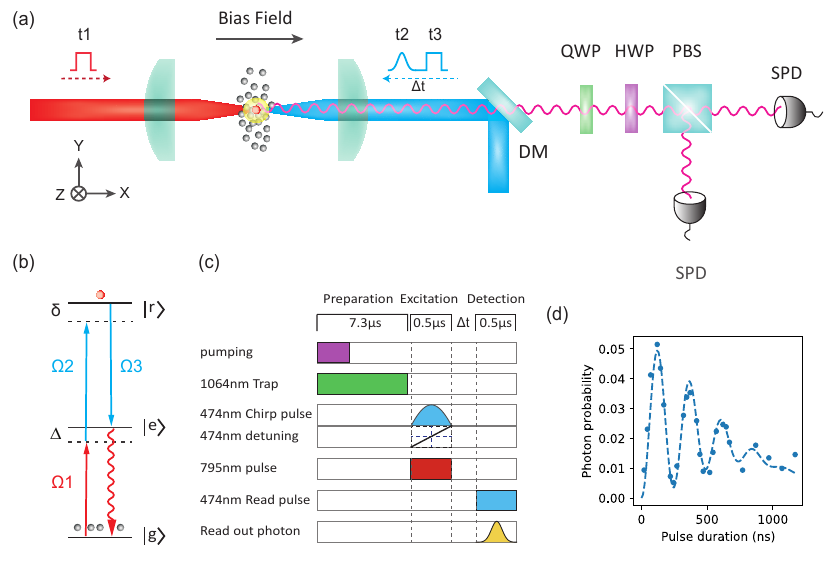}
\caption{(a) Experimental layout. The two excitation lasers ($\Omega_1$ and $\Omega_2$) shine the atomic ensemble in opposite directions ($x$ versus $-x$). A read laser $\Omega_3$ shines through the same direction as $\Omega_2$, leading the retrieval of single photons in the direction of $x$ due to phase matching. A dichroic mirror (DM) is used to couple in the blue lasers and filter the read-out photons. The photons are later rotated with a quarter-waveplate (QWP) and a half-waveplate (HWP) before being split with a polarized beamsplitter (PBS) and detected with single-photon detectors (SPD). (b) Energy levels and laser frequencies. (c) Time sequences of the experiment. (d) Observed collective Rabi oscillations. The sum of detection probabilities is plotted as a function of pulse duration.}
\label{fig1}
\end{figure*}

In our experiment, we make use of laser-cooled Rubidium atoms ($^{87}$Rb), with a temperature of 32~$\upmu$K. An ensemble of atoms are confined with an sliced optical dipole trap (1064~nm), with an effective thickness of 6.3 $\upmu$m. As shown in Fig.~\ref{fig1}, we make use a two-photon scheme to excite atoms from a ground state ($\ket{g}=\ket{5S_{1/2},F=2,m_F=+2}$) to a Rydberg state ($\ket{r}=\ket{81S_{1/2},m_J=+1/2}$), via an intermediate state ($\ket{e}=\ket{5P_{1/2},F=1,m_F=+1}$). One laser at 795 nm couples
with the $|g\rangle \leftrightarrow |e\rangle$ transition, with a beam waist of $w_1=7.4$~$\upmu$m at the atom location and a typical Rabi frequency of $\Omega_1/(2\pi)=2.3$~MHz. A second laser at 474 nm couples with the $|e\rangle \leftrightarrow |r\rangle$ transition, with a beam waist of $w_2=8$~$\upmu$m and a typical Rabi frequency of $\Omega_2/(2\pi)=10.7$~MHz. The single-photon detuning is $\Delta_1/(2\pi)=-40$~MHz for the 795~nm laser. While for the 474~nm excitation laser, the single-photon detuning $\Delta_2$ is close to $-\Delta_1$, being variable. The two-photon excitation has an effective detuning of $\delta = \Delta_1+\Delta_2$, and an effective single-atom Rabi frequency of $\Omega=\Omega_1\Omega_2/(2\Delta_1)$. The excitation lasers select a region with an effective transversal size of $\sqrt{(\pi/2)(1/w_1^2 + 1/w_2^2)^{-1}}=6.8$~$\upmu$m. The collective excitation in the Rydberg state is detected optically by applying a read laser (474 nm) with a Rabi frequency of $\Omega_3$. The read laser converts atomic excitations into single photons in a well-defined mode through phase matching. By using a beam splitter together with two single-photon detectors, we are able to measure the single-photon probability and perform Hanbury-Brown-Twiss~(HBT) test as well. More details on the experimental setup can be found in our previous publications~\cite{li2016hong,Li2019}.

\begin{figure}[t]
\includegraphics[width=\columnwidth]{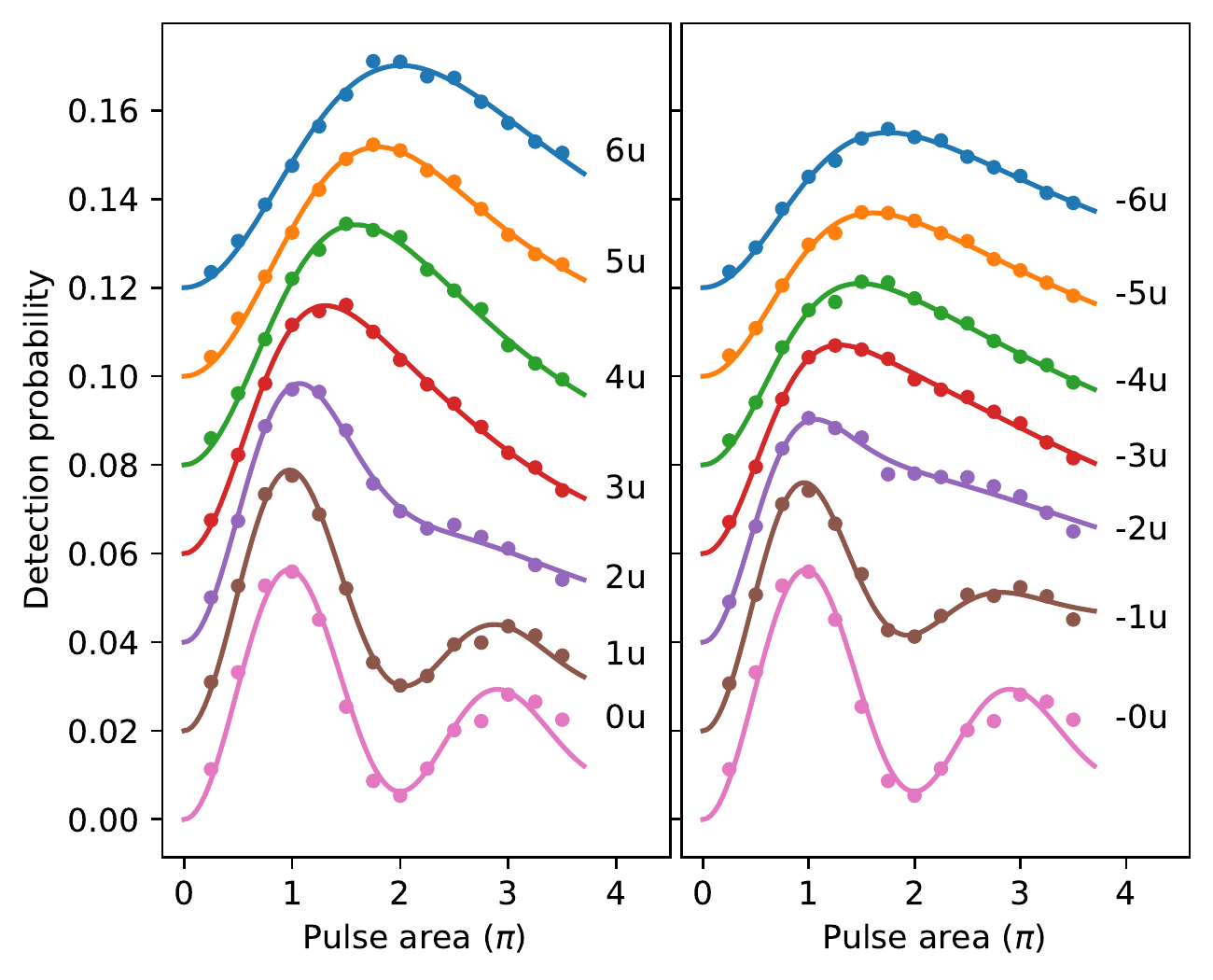}
\caption{Measurement of photon detection probability as a function of pulse area for a series of chirp rates. Result of positive chirping is shown in the left panel, while negative chirping is shown in the right panel. The chirp rate varies with a step of u = $2\pi\times12$~MHz/$\upmu$s. For clarity, data for each chirp rate is shifted vertically with a step of 0.02 in sequence. Each curve is fitted with a function of $y = a \exp(-b\,x^2)[1 - \exp(-c\,x^2) \cos(d\,x)]$ where $a$, $b$, $c$, $d$ are fitting parameters~\cite{dudin2012observation}.}
\label{fig2}
\end{figure}

By applying the two driving lasers, and adiabatically eliminating the intermediate state since $\Omega_1,\Omega_2 \ll \Delta$, we get the eigen states of the system as follows (if the blockade is strong enough)
\begin{equation}
    \begin{cases}
    \ket{+}=\sin\theta\ket{G}+\cos\theta\ket{R}\\
    \ket{-}=\cos\theta\ket{G}-\sin\theta\ket{R },
    \end{cases}
\end{equation}
where $|G\rangle$ refers to the case that all atoms are in the ground state $|g\rangle$, $\ket{R}=N^{-1/2}\sum_{i=1}^{N}\ket{r_i}$ refers to a singly excited state with one atom in the Rydberg state $|r\rangle$ and all other atoms remaining in the ground state $|g\rangle$, and $\theta=\arctan(\Omega_N/\delta)/2$ is the mixing angle with $\Omega_N$ being the collective Rabi frequency. By setting the two-photon detuning to be $\delta=0$, the system eigen states are $|\pm\rangle=|G\rangle\pm|R\rangle$, thus the atomic ensemble undergoes collective Rabi oscillations between the state $|G\rangle$ and $|R\rangle$. A typically Rabi oscillation measured is shown in Fig.~\ref{fig1}(d), from which we can estimate the effective number of atoms in the region of interest is around 180 via the relation of $N = \Omega_N^2 / \Omega^2$~\cite{dudin2012observation,Saffman2010c}. By setting the pulse duration for a pulse area of $\pi$, we get the maximal excitation probability. In the ideal case of full blockade with fixed $N$, $\delta$ and pulse duration, such an excitation process is in principle deterministic.

In the case of chirped excitation~\cite{beterov2011deterministic}, the detuning $\delta$ sweeps across $\delta=0$, and the atoms adiabatically evolves from $|G\rangle$ to $|R\rangle$ (via $|-\rangle$ for positive chirping or $|+\rangle$ for negative chirping), thus deterministic excitation of a single Rydberg atom can be also realized. Such an adiabatic process is robust, being insensitive to $\Omega_N$, $\delta$ and pulse duration. To ensure that the eigen state evolution does not introduce orthogonal components, one has to fulfill the adiabatic condition, $|\dot{\Omega}_N\delta-\Omega_N\dot{\delta}| \ll 2(\Omega_N^2+\delta^2)^{3/2}$~\cite{Vitanov2001}. Thus, it requires a smooth pulse shape. In our experiment, the excitation process has a duration of $2T = 500$~ns. We shape the 474~nm pulse as Gaussian, and shape the 795~nm pulse as squared, which together ensure the smooth evolution of $\Omega$ over time. The full width at half maximum (FWHM) is 188 ns for 474~nm and 467 ns for 795 nm respectively. Frequency chirping is realized via linearly sweeping the 474~nm laser around a central frequency of $\Delta_2 = 40$~MHz with an acousto-optic modulator configured in double-pass.

\begin{figure}[htb]
\includegraphics[width=\columnwidth]{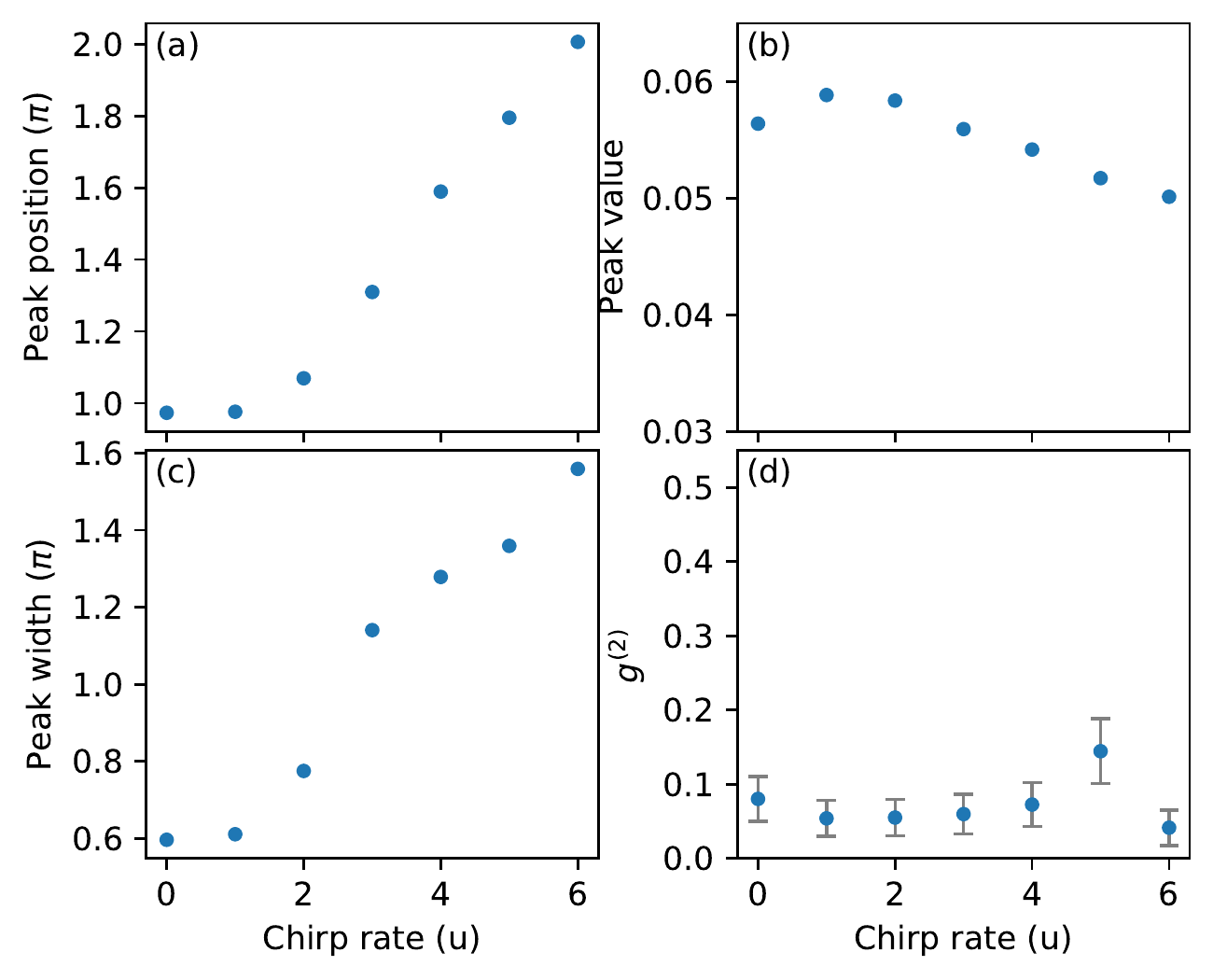}
\caption{For all the positive chirping data shown in Fig.~\ref{fig2}, we summarize the peak positions in (a), the peak values in (b), and the peak widths (defined the region above 80\% of the peak value) in (c). For each chirp rate, we also give a measured $g^{(2)}$ value which is selected for a point which is closest to the peak position in terms of pulse area.}
\label{fig3}
\end{figure}

First we measure the photon detection probability as a function of pulse area for a series of different chirp rates $\alpha$ by changing laser power, with the result shown in Fig.~\ref{fig2}. As the chirp rate increases, the result shows a variation from damped oscillations to damped plateaus, which is a clear signature of transition from Rabi oscillations to adiabatic evolutions. The damping is mainly due to inhomogeneous light shifts induced by the contribution of minor double excitations~\cite{dudin2012observation} since the Rydberg interaction is not strong enough. It is also very interesting to compare the result of positive chirping with negative chirping. Clearly the positive chirping performs much better, which is due to influence of the intermediate state $|e\rangle$. During the chirping process, the detuning $\Delta_2$ relative to the transition $|e\rangle \leftrightarrow |r\rangle$ evolves from $|\Delta_1| - \alpha T$ to $|\Delta_1| + \alpha T$. Thus $|\Delta_2|$ increases gradually as a function of time for positive chirping, while decreases for negative chirping. Considering the gradual increase of Rydberg population, the positive chirping scheme will thus suffer less from the off-resonant coupling and preserves the Rydberg excitation better. In the following, we focus on positive chirping only.

In order to quantitatively evaluate the performance of chirped excitation, we fit the results in Fig.~\ref{fig2} and get peak values and peak positions, which are summarized in Fig.~\ref{fig3}(a, b). For the evaluation of robustness over the pulse area, we define the peak width as the range where photon probability is above 80\% of the peak value. We derive the peak width from the fitted function for each chirp rate, with results summarized in Fig.~\ref{fig3}(c). It is clear to see that the peak width increases significantly as the increase of chirp rate. For a chirp rate of $\alpha = 6$u, the robustness is 2.6 times larger than the $\pi$-pulse excitation ($\alpha = 0$). While, we see that although a higher chirp rate gives rise to better robustness, it requires a larger pulse area than $\pi$. Simultaneously, as the increase of chirp rate, the maximal photon probability achieved gets smaller, which is due to increased contribution of double excitations. In addition, it is very important to verify that the chirping scheme creates genuine single-photons. In our experiment, along with the measurement of photon probability, we perform HBT test of the retrieved optical field simultaneously. We find that the chirping scheme does not influence the single-photon character in comparison with $\pi$-pulse excitation. For reference, we give the measured $g^{(2)}$ values for each chirping rate in Fig.~\ref{fig3}(d). For each $\alpha$, we choose a data point which is closest to the peaks in Fig.~\ref{fig2}. We see that the variance is within the range of error bars.

\begin{figure}[htb]
\includegraphics[width=1\columnwidth]{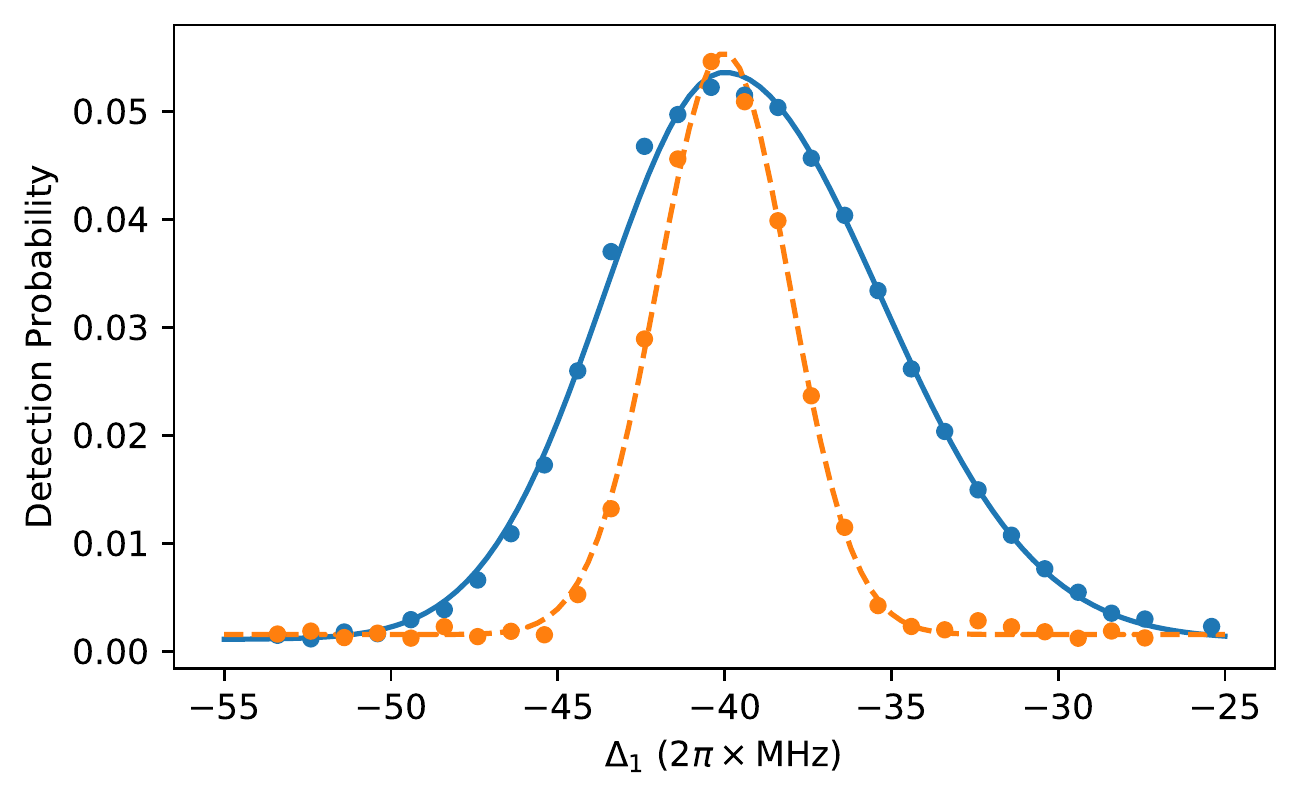}
\caption{Measurement of photon detection probability as a function of single-photon detuning $\Delta_1$. Result of chirping scheme is shown as solid line in blue, while result of $\pi$-pulse scheme is shown as dashed line in orange. Data points are fitted with asymmetric Gaussian functions. The two results are measured with the same laser powers. The pulse width (FWHM) determined by the 474~nm laser is 188~ns for the chirping scheme and 114~ns for the $\pi$-pulse scheme respectively. For the chirping scheme, we set a chirp rate of $\alpha = 4$u. The pulse area is optimized to maximize the photon probability at $\Delta_1 = -40$~MHz for the two schemes respectively.}
\label{fig4}
\end{figure}

The excitation scheme with chirped pulses is not only robust against pulse area but also against other parameters, such as the laser frequency detunings. In our experiment, by setting the best pulse area, we deliberately shift $\Delta_1$ and test the robustness of photon probability against it. The result is shown in Fig.~\ref{fig4}, in which we compare the chirping scheme with the $\pi$-pulse scheme. It is clear to see that, the chirping scheme is much robust. Similar as the case in Fig.~\ref{fig2}, we also define a robustness range that the photon probability is higher than 80\% of the peak value. We find that the range of robustness for chirped excitation is 2.1 times larger than $\pi$-pulse excitation. In the case of chirped excitation, we also find obvious asymmetry along the axis $\Delta_1=-40$~MHz, which we think is due to the increase of single-atom Rabi frequency $|\Omega|$ as the reducing of $|\Delta_1|$.

To summarize, we realize the creation of single Rydberg excitations with chirped pulses via adiabatic rapid passage. The created excitations are verified to be phase coherent via directional retrieval as optical fields. The quantum signature of the excitations are tested via photonic HBT test, which suggests that genuine single excitations are created and can be further used for quantum information applications.
The detected single-photon probability is similar as traditional $\pi$-pulse excitation. We also test the robustness against variance of pulse area and two-photon detuning, both results suggest that the chirping scheme is more robust than $\pi$-pulse excitation. Therefore, our work proves the validity of applying the technique of chirped excitation for an ensemble system under Rydberg blockade, and may become an elementary tool in future experiments of quantum computing with collective encoding~\cite{Brion2007,Saffman2008,li2016hong}, and quantum network with Rydberg super atoms~\cite{Han2010,Zhao2010c,Li2013g, Li2019}.

This work was supported by National Key R\&D Program of China (No. 2017YFA0303902), Anhui Initiative in Quantum Information Technologies, National Natural Science Foundation of China, and the Chinese Academy of Sciences. J. L. acknowledges support from China Postdoctoral Science Foundation (No. 2017M622000).

\bibliography{refs}

\end{document}